\documentclass[iop, apj, revtex4]{emulateapj}
\usepackage{natbib}
\usepackage{url}
\usepackage{amsmath}
\usepackage{appendix}
\usepackage{hyperref}

\newcommand{\kms}{km s$^{-1}$}

\begin{document}

\title{NOEMA Observations of a Molecular Cloud in the low-metallicity Galaxy Kiso
5639}

\author{Bruce G. Elmegreen\altaffilmark{1}, Cinthya Herrera\altaffilmark{2},
Monica Rubio\altaffilmark{3}, Debra Meloy Elmegreen\altaffilmark{4}, Jorge
S\'anchez Almeida\altaffilmark{5}, Casiana Mu\~noz-Tu\~n\'on\altaffilmark{5},
Amanda Olmo-Garc\'ia\altaffilmark{5}}

\altaffiltext{1}{IBM Research Division, T.J. Watson Research Center, 1101 Kitchawan
Road, Yorktown Heights, NY 10598; bge@us.ibm.com}

\altaffiltext{2}{Institut de Radioastronomie Millim\'etrique, 300 rue de la
Piscine, Domaine Universitaire, F-38406, Saint-Martin-d'H\`eres, France}

\altaffiltext{3}{Departamento de Astronomia, Universidad de Chile, Casilla 36-D,
Santiago, Chile}

\altaffiltext{4}{Department of Physics \& Astronomy, Vassar College, Poughkeepsie, NY
12604}

\altaffiltext{5}{Instituto de Astrof\'isica de Canarias, C/ via L\'actea, s/n, 38205,
La Laguna, Tenerife, Spain, and Departamento de Astrof\'isica, Universidad de La
Laguna}

\begin{abstract}
A giant star-forming region in a metal-poor dwarf galaxy has been observed in
optical lines with the 10-m Gran Telescopio Canarias and in the emission line of
CO(1-0) with the NOEMA mm-wave interferometer. The metallicity was determined to
be $12+\log({\rm O/H})=7.83\pm 0.09$, from which we estimate a conversion factor
of $\alpha_{\rm CO}\sim100\;M_\odot\;{\rm pc}^{-2}\left( \rm{K \;km\;
s}^{-1}\right)^{-1}$ and a molecular cloud mass of $\sim2.9\times10^7\;M_\odot$.
This is an enormous concentration of molecular mass at one end of a small
galaxy, suggesting a recent accretion. The molecular cloud properties seem
normal: the surface density, $120\;M_\odot$ pc$^{-2}$, is comparable to that of
a standard giant molecular cloud, the cloud's virial ratio of $\sim1.8$ is in
the star-formation range, and the gas consumption time, $0.5$ Gyr, at the
present star formation rate is typical for molecular regions. The low
metallicity implies that the cloud has an average visual extinction of only
$0.8$ mag, which is close to the threshold for molecule formation. With such an
extinction threshold, molecular clouds in metal-poor regions should have high
surface densities and high internal pressures. If high pressure is associated
with the formation of massive clusters, then metal-poor galaxies such as dwarfs
in the early universe could have been the hosts of metal-poor globular clusters.
\end{abstract}

\section{Introduction}
\label{intro}

Kiso 5639 \citep{kiso-all} is a dwarf galaxy with a kpc-size starburst at one end,
giving the system a tadpole or cometary shape \citep{elm12}.  The rotation speed
of $\sim 35$ km s$^{-1}$ \citep{jorge13} combined with a radius of 1.2 kpc in the
bright part of the disk implies that the dynamical mass there is $3\times
10^8/\sin^2 i\;M_{\odot}$, which is a factor of $\sim 6$ larger than the stellar
mass of $5\times 10^7\;M_{\odot}$ from Sloan Digital Sky Survey photometry
\citep{elm12} and comparable to the total H\textsc{i} mass of
$\sim3\times10^8\;M_\odot$ \citep{salzer}.

The galaxy is a member of our spectroscopic survey of 22 low metallicity dwarfs
where the metallicity in the starburst ``head'' appears to be less than in the
rest of the galaxy (the ``tail''); 16 others in this survey have the same
metallicity drop, 3 do not and 2 are ambiguous \citep{jorge13,jorge14,jorge15}.
This peculiar pattern of metallicity suggests that the starbursts in these systems
were triggered by accreting gas with lower metallicity than in the rest of the
galaxy. Other examples of metallicity drops were reported in \cite{levesque11},
\cite{haurberg13} and \cite{lagos18}.

HST observations of Kiso 5639 \citep{elmegreen16} in six UV-optical and H$\alpha$
filters were used to resolve the head and derive the star formation properties.
The head contains 14 young star clusters more massive than $10^4\;M_\odot$ and an
overall clustering fraction for star formation of $25-40$\%. The H$\alpha$
luminosity of the core region of the head is $8.8\pm0.16\times10^{39}$ erg
s$^{-1}$ inside an area of $3.6\times3.6$ square arcsec. The corresponding star
formation rate is $\sim0.04\;M_{\odot}$ yr$^{-1}$. This rate is based on a
conversion factor of $SFR=4.7\times10^{-42}L(H\alpha)\;M_\odot$ yr$^{-1}({\rm
erg\;s}^{-1})^{-1}$ that is appropriate for low metallicity, which introduces a
factor of $0.87$ times the standard value \citep{hunter10,kennicutt12}. For a
distance of 24.5 Mpc \citep{elm12}, the corresponding area is $0.18$ kpc$^2$
($430\times430$ square pc), and the star formation rate per unit area is
$0.23\;M_\odot$ pc$^{-2}$ Myr $^{-1}$.

This is a high rate for a small galaxy and it suggests there is a large reservoir
of dense gas in the head.  For the conventional molecular gas consumption time of
$\sim2$ Gyr \citep{bigiel08}, the molecular mass surface density would be
$450\;M_\odot$ pc$^{-2}$, which is $\sim40$ times higher than the average stellar
surface density and $\sim7$ times higher than the average dynamical surface
density in the bright part of the disk. For the Kennicutt-Schmidt relation in
whole galaxies from Figure 11 in \cite{kennicutt12}, the required gas surface
density would be $\sim200\;M_\odot$ pc$^{-2}$. These high gas surface densities
suggest that Kiso 5639 is lopsided and recently accreted at least
$\sim10^8\;M_\odot$ of gas on one side. Simulations of such a process are in
\cite{verbeke14} and \cite{ceverino16}.

Detection of CO emission from the head of Kiso 5639 would help to clarify the
situation. The metallicity of this galaxy, determined with the 2.5m Nordic Optical
Telescope from the [NII]$\lambda6583$ to H$\alpha$ line ratio in \cite{jorge13},
was $12+\log({\rm O/H})\sim7.48\pm0.04$. CO is highly underabundant compared to
H$_2$ at low metallicity \citep{leroy11}. The lowest metallicities in regions
detected in CO, i.e., $7.5$ in Sextens A and DDO 70 \citep{shi15,shi16} and 7.8 in
WLM \citep{elmegreen13,rubio15} occur in dwarf Irregular galaxies where the CO
clouds are small and in the centers of large HI and H$_2$ clouds. For the large
molecular mass expected in Kiso 5639 and in other galaxies of its type, CO
emission should be more evident.

To search for CO in Kiso 5639, we observed CO(1-0) in the head region with NOEMA
in D configuration. The result was a detection corresponding to a gas surface
density similar to that estimated above, depending on the assumed $\alpha_{\rm
CO}$ conversion factor. Because the metallicity is important for this $\alpha_{\rm
CO}$, we re-observed Kiso 5639 in several optical lines including
[OII]$\lambda$3727 and [OIII]$\lambda$4363 using the 10m Gran Telescopio Canarias
(GTC) to get the metallicity again. The observations, results, and implications
are discussed in the next three sections.

\section{Observations}
\subsection{Metallicity}
The metallicity of Kiso 5639 is central to the discussion on the origin of the CO
emission. \cite{jorge13} found a drop in metallicity coinciding with its largest
star-forming region, with the value of $12+\log({\rm O/H})\simeq 7.5$. This
estimate was based on the strong-line ratio N2 ($\log([{\rm NII}]\lambda 6583/{\rm
H}\alpha$)), which is known to have biases \citep[e.g.,][]{morales14}. On the
other hand, the SDSS spectrum of the region gives a metallicity of around 7.8 when
using either the direct method (DM) or the HIICM procedure by Perez-Montero (2014)
(Ruben Garc\'\i a-Benito 2017, private communication). In order to remove
uncertainties and secure the metallicity estimate, we obtained long-slit spectra
of the galaxy integrating for 2.5 hours with the instrument OSIRIS at the 10-m GTC
telescope. The slit was placed along the major axis.

The resulting visible spectra cover from 3600 \AA\ to 7200 \AA\ with a spectral
resolution around 550, thus containing all the spectral lines listed below that
are needed for the metallicity measurement, including the critically important
temperature sensitive line [OIII]$\lambda$4363. The data were calibrated in flux
and wavelength using PyRAF\footnote{{\tt
http://www.stsci.edu/institute/software$\_$hardware/pyraf}}, and Gaussian fits
provided the fluxes of the emission lines. We employed the code HIICM to infer
O/H, which requires  [OII]$\lambda$3727, [OIII]$\lambda$4363, H$\beta$,
[OIII]$\lambda$5007, H$\alpha$, [NII]$\lambda$6583, and
[SII]$\lambda\lambda$6717,6737. HIICM was chosen because it is robust and
equivalent to the direct method when [OIII]$\lambda$4363 is available
\citep{perez14,sanchez16}. The oxygen abundance thus obtained was $12+\log({\rm
O/H}) = 7.83 \pm 0.09$, considering $4^{\prime\prime}$ around the star-forming
region. HIICM estimates error bars from the difference between the observed line
fluxes and those predicted by the best fitting photoionization model. The inferred
metallicity is similar when all the spectra of the galaxy are integrated ($7.78
\pm 0.11$), and even if only spectra from the faint tail are considered ($7.79 \pm
0.18$).  HIICM also provides N/O, which turns out to be roughly constant with a
value around $\log({\rm N/O})\simeq -1.5\pm 0.1$, typical for the metallicity
assigned to Kiso 5639 \citep[e.g.,][]{vincenzo16}.

Curiously, the new data still show a slight metallicity drop at the head from the
N2 method, down to $12+\log({\rm O/H}) \sim 7.6$, with the same value, $\sim7.8$,
as for the HIICM measurements elsewhere. There is also a slight indication of a
drop in N/O in the head, by a factor of $\sim3$, which is a $\sim3\sigma$
deviation for the head measurement compared to the average in the tail, but only a
$1\sigma$ devitation for the weaker tail measurement in comparison to the head.
This apparent drop in head metallicity from the [NII] line explains the
measurement in \cite{jorge13}, but it is not viewed as relevant for our CO
observations, which need the Oxygen abundance. Thus we use the HIICM value of
$12+\log({\rm O/H}) = 7.83 \pm 0.09$ to determine $\alpha_{\rm CO}$ in the head
where CO is detected.

\subsection{CO Observations}

We obtained observations of Kiso 5639 on May, July and August 2017, with the
Northern Extended Millimeter Array (NOEMA) at Plateau de Bure, in the CO ($J=1-0$)
line, at 114.572~GHz redshifted to the velocity of Kiso 5639. The observations
were carried out with either 7 or 8 antennas in Configuration D, with baselines
between 15~m and 175~m. We used the Widex correlator, with a total bandwidth of
3.6~GHz and a native spectral resolution of 1.95~MHz (5.1~km s$^{-1}$).

Data reduction, calibration and imaging were performed with CLIC and MAPPING
softwares of GILDAS\footnote{http://www.iram.fr/IRAMFR/GILDAS/}, using standard
procedures. Images were reconstructed using the natural weighting, resulting in a
synthesized beam of 2$\farcs$8$\times$3$\farcs$7 (PA=26$^{\circ}$).   The rms
noise in the CO cube is 14.3 mK (1.6 mJy beam$^{-1}$) in a 5.1~km s$^{-1}$
channel.

The resulting data cube was integrated between LSR velocities of $-62.3$ km
s$^{-1}$  and $-26.5$ km s$^{-1}$ where CO emission was observed. The rms was
measured in the integrated image and a mask for pixels with emission greater than
$ 2\sigma$ was made. The average spectrum of Kiso 5639 was obtained within this
mask. A 1D Gaussian fit to this spectrum gives $V_{\rm LSR}= -42.1\pm 2.2$ km
s$^{-1}$, FWHM$=35.2\pm5.3$ km s$^{-1}$, peak emission $32.1\pm4.5$ mK, and
average integrated line profile, $1203\pm247$ mK km s$^{-1}$, as summarized in
Table 1.

\begin{figure*}
\epsscale{1.} \plotone{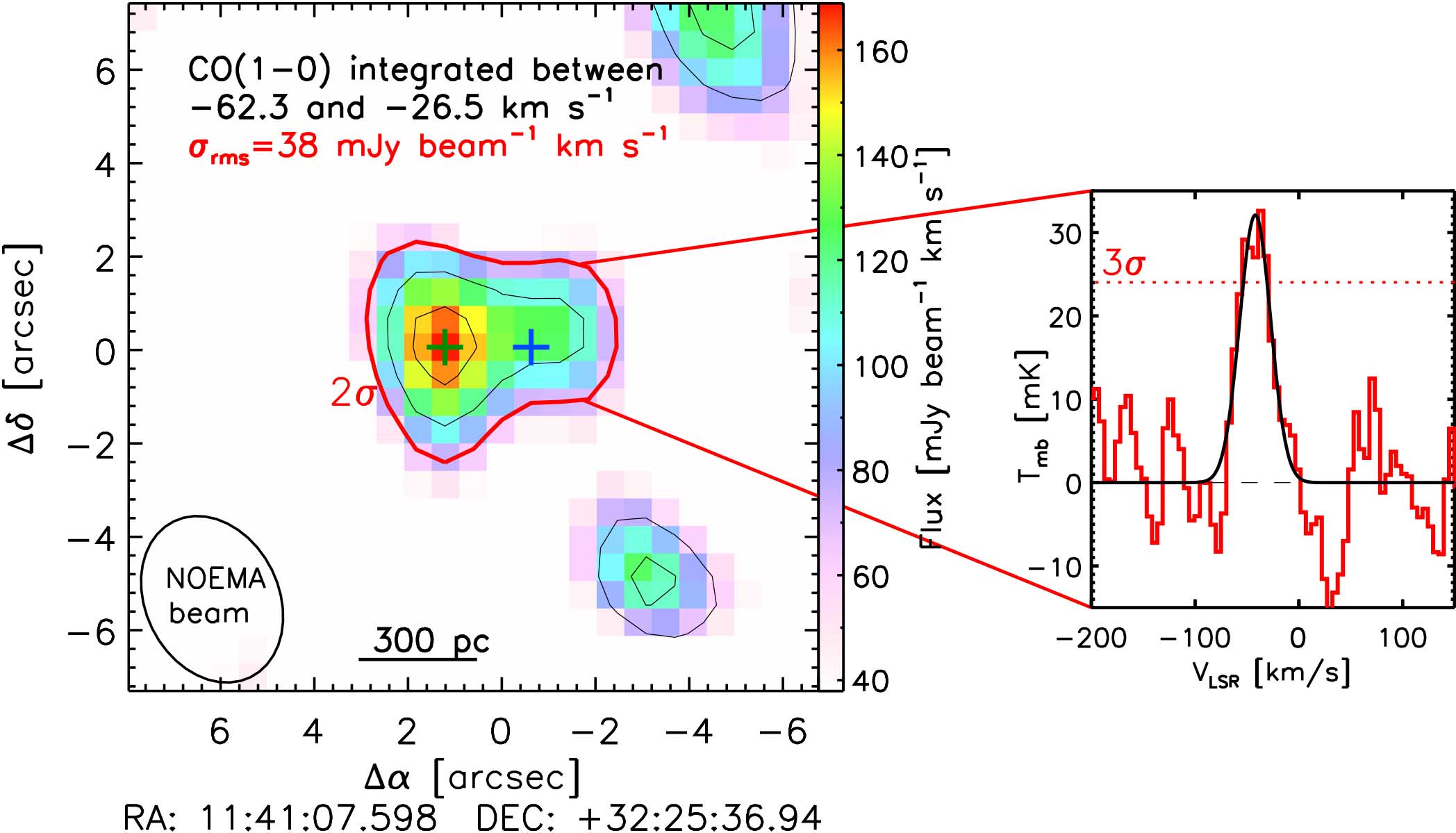}
\caption{
Left panel: color image shows the  CO(1$-$0) emission integrated from $-62.3$
to $-26.5$  \kms\ in units of Jy beam$^{-1}$ \kms. The blue plus symbols are the
centers of the two velocity components. The size of the beam of
$2\farcs$8$\times$3$\farcs7$ is shown in the bottom-left corner. Offset positions are
relative to $\alpha$: 11$^{\rm h}$41$^{\rm m}$7$\fs$598, $\delta$: 32$^{\circ}$25$\arcmin$36$\farcs$94
J2000. Contours correspond to the emission at 2$\sigma$, 3$\sigma$ and 4$\sigma$,
with $\sigma=$38 mJy beam$^{-1}$ \kms. The red contour highlights the 2$\sigma$
emission. Right panel: CO(1$-$0) line profile integrated over the area defined
by the 2$\sigma$ contour, in temperature brightness. Black line is the fit
to the spectrum using a Gaussian curve. The 3$\sigma$ level
of the spectrum is marked.} \label{Fig1_CH_23apr2018}
\end{figure*}

\begin{figure*}
\epsscale{1.} \plotone{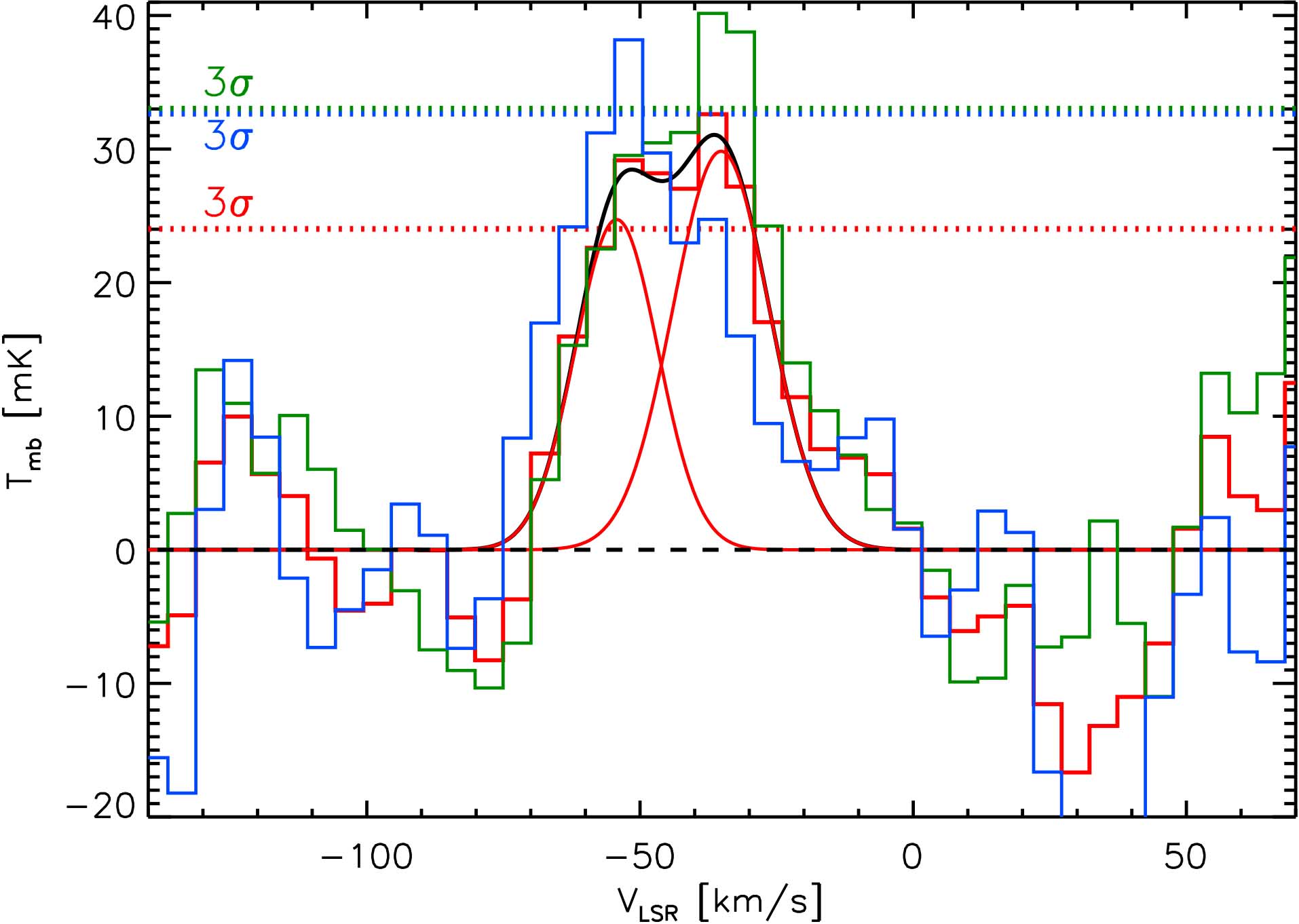}
\caption{CO(1$-$0) spectra for the different cloud components. The red histogram is
the CO(1$-$0)
line profile for the main cloud, as shown in Fig.~\ref{Fig1_CH_23apr2018}. The
blue and green histograms
are obtained by integrating at the positions of the blue and green crosses
in Fig.~\ref{Fig1_CH_23apr2018}, within an aperture size equal to
the beam size. The red curves are Gaussian fits to the individual components
and the black curve is the fit to the
main cloud spectrum using two components. We have marked the 3$\sigma$
levels for each spectrum.} \label{Fig2_CH_23apr2018}
\end{figure*}

Figure \ref{Fig1_CH_23apr2018} shows a map of the main emission and the integrated
spectrum. The velocity resolution is 5.11 km s$^{-1}$, the half-power beam size is
$2.85^{\prime\prime}\times3.72^{\prime\prime}$, and the rms is $1.6$ mJy/beam. The
equivalent radius of the source is $2.35^{\prime\prime}$ (280 pc), which
corresponds to the area of the detection contour, $2.45\times10^5$ pc$^{2}$. The
emission has two peaks separated by $2^{\prime\prime}$, which is 240 pc. Spectra
determined for each peak are shown in Figure \ref{Fig2_CH_23apr2018}.   The blue
histogram is from one of the peaks and the green histogram is from the other; the
red double peak histogram is the composite. Each histogram has a Gaussian fit, as
indicated by the red lines and the black line. The integrated line fits to these
spectra and the derived quantities are in Table 1.

Several other regions inside the primary beam of the NOEMA telescope were also
suspected of containing CO emission, but the brightness temperatures were less
than 20\% of the peak shown in the figure, and we do not consider them to be
definitive detections of molecular gas.  These include the contoured regions to
the southwest and northwest of the main emission in Figure
\ref{Fig1_CH_23apr2018}.

\section{Results}

Conversion from the observed CO flux to a molecular mass depends on the conversion
factor, $\alpha_{\rm CO}$. \cite{hunt15} suggested an extrapolation of
$\alpha_{\rm CO}$ from the solar neighborhood value to low metallicity $Z$ as
$\alpha_{\rm CO}=4.3(Z/Z_\odot)^{-2}\;M_\odot\;{\rm pc}^{-2}\left( \rm{K \;km\;
s}^{-1}\right)^{-1}$, where $Z_\odot$ is the solar metallicity corresponding to
$12+\log({\rm O/H}) = 8.69$ \citep{asplund09}, and $Z$ is the metallicity for Kiso
5639 corresponding to $12+\log({\rm O/H}) = 7.83$. The local conversion factor is
$\alpha_{\rm CO,\odot}=4.3 \;M_\odot\;{\rm pc}^{-2}\left( \rm{K \;km\;
s}^{-1}\right)^{-1}$ including He and heavy elements \citep{bolatto13}. Similarly,
\cite{amorin16} suggested an extrapolation as $Z$ to the power $-1.5$. In the
first case, the result for Kiso 5639 would be $\alpha_{\rm CO}=225$ in these
units, and in the second case it would be $84$. Compared to the WLM galaxy, where
$\alpha_{\rm WLM}\sim124\pm60$ \citep{elmegreen13} and $12+\log({\rm O/H})_{\rm
WLM}=7.8$, $\alpha_{\rm CO}=108$ and 112 for the same power law scalings. On the
other hand, if we extrapolate between the low-$Z$ $\alpha_{\rm CO}$ values
determined by \cite{shi16}, we would get $\alpha_{\rm CO}\sim600$ and 900 at
$Z=7.8$, respectively, which results in a much larger molecular mass. Here we take
$\alpha_{\rm CO}\sim100$ as a conservative estimate.

For the observed CO average integrated line profile, $1.2\pm0.2$ K km s$^{-1}$,
estimated $\alpha_{\rm CO}=100\;M_\odot\;{\rm pc}^{-2}\left( \rm{K \;km\;
s}^{-1}\right)^{-1}$, and CO source area, $2.45\times10^5$ pc$^2$, the H$_2$ mass
is the product of these, $2.9\times10^7\;M_\odot$, with the largest uncertainty in
the value of $\alpha_{\rm CO}$, which is probably a factor of 2. The mass divided
by the total emitting area is the molecular surface density,  $\Sigma_{\rm mol}=
120\;M_\odot$ pc$^{-2}$.

In this same region, the star formation rate is $\sim0.04\;M_{\odot}$ yr$^{-1}$
and the star formation surface density is $0.23\;M_\odot$ pc$^{-2}$ Myr $^{-1}$,
as given above.  These imply that the molecular gas consumption time is $0.5-0.7$
Gyr, which is slightly less than in normal spiral galaxy disks
\citep{bigiel08,schruba11}.   This time is about the same as in the SMC, where the
metallicity is also low ($Z\sim0.2Z_\odot$) and $\alpha_{\rm CO}$ is high,
$\sim220\;M_\odot\;{\rm pc}^{-2}\left(\rm{K \;km\; s}^{-1}\right)^{-1}$
\citep{bolatto11}.

Following the same method, the masses of the two components of the emission in
Figures \ref{Fig1_CH_23apr2018} and \ref{Fig2_CH_23apr2018} can be determined (see
Table 1). The component in the east has a velocity of $-35$ km s$^{-1}$ and an
average integrated line profile of $0.66\pm0.25$ K km s$^{-1}$, and the component
in the west has a velocity of $-54$ km s$^{1}$ and an average integrated line
profile of $0.47\pm0.22$ K km s$^{-1}$. We assume each has the same total area.
Then the masses are these line integrals multiplied by the areas and the
$\alpha_{\rm CO}$ factor, which are $1.6\times10^7 \;M_\odot$ in the east and
$1.1\times10^7 \;M_\odot$ in the west, with surface densities of $66\;M_\odot$
pc$^{-2}$ and $47\;M_\odot$ pc$^{-2}$, respectively.  The velocity difference
between the components, $19$ km s$^{-1}$, is nearly half of the rotation velocity
of the galaxy, $34.7\pm6.2$ km s$^{-1}$ \citep{jorge13}, suggesting a catastrophic
event.

The masses, radii, and velocity dispersions can be combined to determine the
virial ratio, $5R\sigma^2/(GM)$, where $\sigma=0.42\times{\rm FWHM}$. These ratios
require a deconvolved radius, $R$ which is
$(2.35^2-2.85\times3.73/4)^{0.5}=1.69^{\prime\prime}$ for the total cloud (using
the source and beam sizes from above). The radii of the components are taken to
have upper limits equal to the average beam size of $1.63^{\prime\prime}$.   The
results are in Table 1. The virial ratios are in the range $<1.1$ to 1.8, which
suggests that the whole cloud and the two separate parts are gravitationally
self-bound.

Figure 3 shows the HST image of Kiso 5639 in H$\alpha$, V, and B bands with
superimposed CO contours at values of $2\sigma$, $3\sigma$ and $4\sigma$. The main
molecular cloud is $2.2^{\prime\prime}=260$ pc to the east of the bright star
formation region. This offset means that comparisons to the Kennicutt-Schmidt
relation are not exactly appropriate. Still, the molecular gas reservoir for the
starburst appears to be of sufficient mass to explain the observed young stars.

Figure 4 shows the same CO contours on the HST H$\alpha$ image of the whole
galaxy. An expanded view of the head region is on the left, where the black
contours are 3\%, 10\% and 20\% of the peak H$\alpha$ emission. Offset positions
are as in Fig.~\ref{Fig1_CH_23apr2018}.

\section{Implications}

Subject to uncertainties about the conversion factor between CO emission and
molecular gas mass, our observations suggest that a $2.9\times10^7\;M_\odot$
molecular cloud with a surface density of $\sim 120\;M_\odot$ pc$^{-2}$ spanning a
region 560 pc in diameter is associated with the starburst head of the tadpole
galaxy Kiso 5639, where HST observations previously suggested that the star
formation rate is $0.04\;M_\odot$ yr$^{-1}$ and the star formation surface density
is $0.23\;M_\odot$ pc$^{-2}$ Myr$^{-1}$.

The molecular surface density is comparable to that in giant molecular clouds in
local galaxies, which average $100-200\;M_\odot$ pc$^{-2}$ \citep{heyer15}, but
the extinction through the molecular cloud in Kiso 5639 should be much less.
Considering the local conversion factor between color excess and HI column density
\citep{bohlin78}, $A_{\rm V}=N/(1.87\times10^{21}\;{\rm cm}^{-2})$, for a ratio of
total to selective extinction $R=3.1$, and using a mean molecular weight of 1.36
times the hydrogen mass, the extinction through a cloud is related to its surface
density by
\begin{equation}
\Sigma_{\rm gas}=20.2A_{\rm V}\left(Z_\odot/Z\right)\;M_\odot\;{\rm pc}^{-2}.
\end{equation}
For Kiso 5639, $Z_\odot/Z=7.2$, so the observed $\Sigma_{\rm
gas}=120\;M_\odot\;{\rm pc}^{-2}$ corresponds to $A_{\rm V}=0.8$ mag. In the solar
neighborhood, this is comparable to the threshold for CO formation
\citep{pineda08,glover12}. The same extinction threshold was obtained for the WLM
galaxy at a metallicity of $12+\log({\rm O/H})=7.8$, where pc-size CO clouds in
the core of a giant HI and H$_2$ envelope had a total shielding column density
equivalent to $\sim1.5$ mag visual extinction (\citealt{rubio15}, see also
\citealt{schruba17}).

The assumed value of $\alpha_{\rm CO} = 100 \;M_\odot \;{\rm pc}^{-2} \left(\rm{K
\;km\; s}^{-1}\right)^{-1}$ implies that the ratio of invisible molecular hydrogen
to observed CO is high, $\sim23$ times higher than in the solar neighborhood. A
similar situation arises for the low-metallicity galaxy WLM, where the CO clouds
resolved by ALMA are pc-scale inside resolved HI and dust clouds that are
$\sim200$ pc in size. If the peak CO antenna temperature of $\sim30$ mK in Kiso
5639 corresponds to a beam-diluted thermal temperature of $\sim30$ K, which is not
unreasonable for a molecular region of intense star formation
\citep[e.g.,][]{glover12}, then the beam-dilution factor of the total CO is
$\sim10^{-3}$, and their individual radii would be $\sim0.03/\sqrt{N}$ of the
overall cloud radius, or $\sim9/\sqrt{N}$ pc for $N$ cores.

Such small CO cores would presumably have a collective motion that is observed as
the CO emission Gaussian linewidth of $\sigma=15$ km s$^{-1}$. Considering again a
typical dense cloud temperature of 30 K, these motions would have a Mach number of
$\sim46$ and a compression ratio in the shocked regions of approximately the
square of this, $\sim2000$. The average compressed density is then the compression
ratio times the average density. The average density is the surface density
divided by the cloud thickness. For a self-gravitating slab, the thickness is
$2\sigma^2/(\pi G \Sigma_{\rm gas})$, which is 140 pc in our case. Then the
average density is $\sim7.6$ cm$^{-3}$. If this average density is compressed by
shocks, then the density in the compressed regions, where the CO might actually be
located, is $1.6\times10^4$ cm$^{-3}$. For a temperature of $\sim30$ K, the
thermal pressure would be $4.8^\times10^5k_{\rm B}$. This derivation of pressure
is the same as what we would get from the equation for self-gravitational binding
pressure, $(\pi/2)G\Sigma_{\rm gas}^2$, using $\Sigma_{\rm gas}\sim120\;M_\odot$
pc$^{-2}$. Any additional contribution to the surface density from atomic gas in
an envelope around the molecular cloud would increase the pressure.

The derived pressure inside the Kiso 5639 molecular cloud is comparable to the
pressures in local giant molecular clouds, corresponding to similar surface
densities.  Pressure has been proposed to control the fraction of mass in the form
of bound clusters in the star formation process \citep{kruijssen12}. Kiso 5639 has
a relatively high clustering fraction, 30\%-45\% \citep{elmegreen16} compared to
other dwarf galaxies \citep{billett02}, perhaps because of its relatively high
pressure. Its clustering fraction is about the same as in spirals \citep{adamo15}
where the pressure is the same.

High pressure might also be necessary to form more massive clusters
\citep{elmegreen01}. The present observation of molecular clouds close to the
extinction threshold for molecule formation \citep[and in][]{rubio15,schruba17}
suggest that the surface densities (and therefore pressures) of star-forming
clouds will be higher at lower metallicity. That is, low metallicity and the
corresponding low dust opacity per column of gas could diminish the ability of
self-gravitating clouds to shield themselves against background starlight. This
would delay the formation of molecules and cold thermal temperatures until the
cloud surface density is high. Then the pressure, which depends only on the
surface density, would become high too.  The result would be a greater likelihood
of gravitationally bound massive clusters in low metallicity galaxies. Massive
concentrations of molecular gas as in Kiso 5639 would be needed too.

We thank R. Garc\'\i a-Benito for deriving the abundance of Oxygen from the SDSS
spectrum. We are grateful to the referee for comments. M.R. wishes to acknowledge
support from CONICYT(CHILE) through FONDECYT grant No1140839 and partial support
from project BASAL PFB-06. M.R. is a member of UMI-FCA, CNRS/INSU, France (UMI
3386). AOG thanks Fundaci\'on La Caixa for financial support in the form of a PhD
contract, and SA and CMT acknowledge MINECO for funding through the project
AYA2016-79724-C4-2-P. This work is based on observations carried out under project
number S17AP with the IRAM NOEMA Interferometer. IRAM is supported by INSU/CNRS
(France), MPG (Germany) and IGN (Spain). Based on observations made with GTC, in
the Spanish Observatorio del Roque de los Muchachos of the IAC, under DDT.

\begin{deluxetable}{lcccccccc}
\tabletypesize{\scriptsize}\tablecolumns{8} \tablewidth{0pt}
\tablecaption{CO cloud Properties}
\tablehead{
\colhead{Cloud}&
\colhead{$V_{\rm LSR}$}&
\colhead{$T_{\rm peak}$}&
\colhead{FWHM} &
\colhead{Ave. Line Int.} &
\colhead{Radius} &
\colhead{Mass}&
\colhead{Surface Density}&
\colhead{Virial Ratio}\\
\colhead{}&
\colhead{km s$^{-1}$}&
\colhead{mK}&
\colhead{km s$^{-1}$}&
\colhead{K km s$^{-1}$}&
\colhead{arcsec}&
\colhead{$\times10^6\;M_\odot$}&
\colhead{$M_{\odot} {\rm pc}^{-2}$}&
\colhead{$5R\sigma^2/(GM)$}
}

\startdata

Total&$-42.1\pm2.2$	&$32.1\pm4.5$	&	$35.2\pm5.3$	&$1203\pm247$ & $1.7$ & $29$ & $120$ & 1.8\\
East&$-35.2\pm3.8$	&$29.8\pm4.3$	&	$20.8\pm7.1$	&$659\pm245$ & $<1.6$ & $16$ & $66$ & $<1.1$\\
West&$-54.3\pm3.8$	&$24.7\pm7.3$	&	$17.8\pm6.5$	&$469\pm219$ & $<1.6$ & $11$ & $47$ & $<1.1$\\

\enddata
\label{tab}
\end{deluxetable}

\newpage

\newpage

\newpage
\begin{figure*}
\epsscale{1.15} \plotone{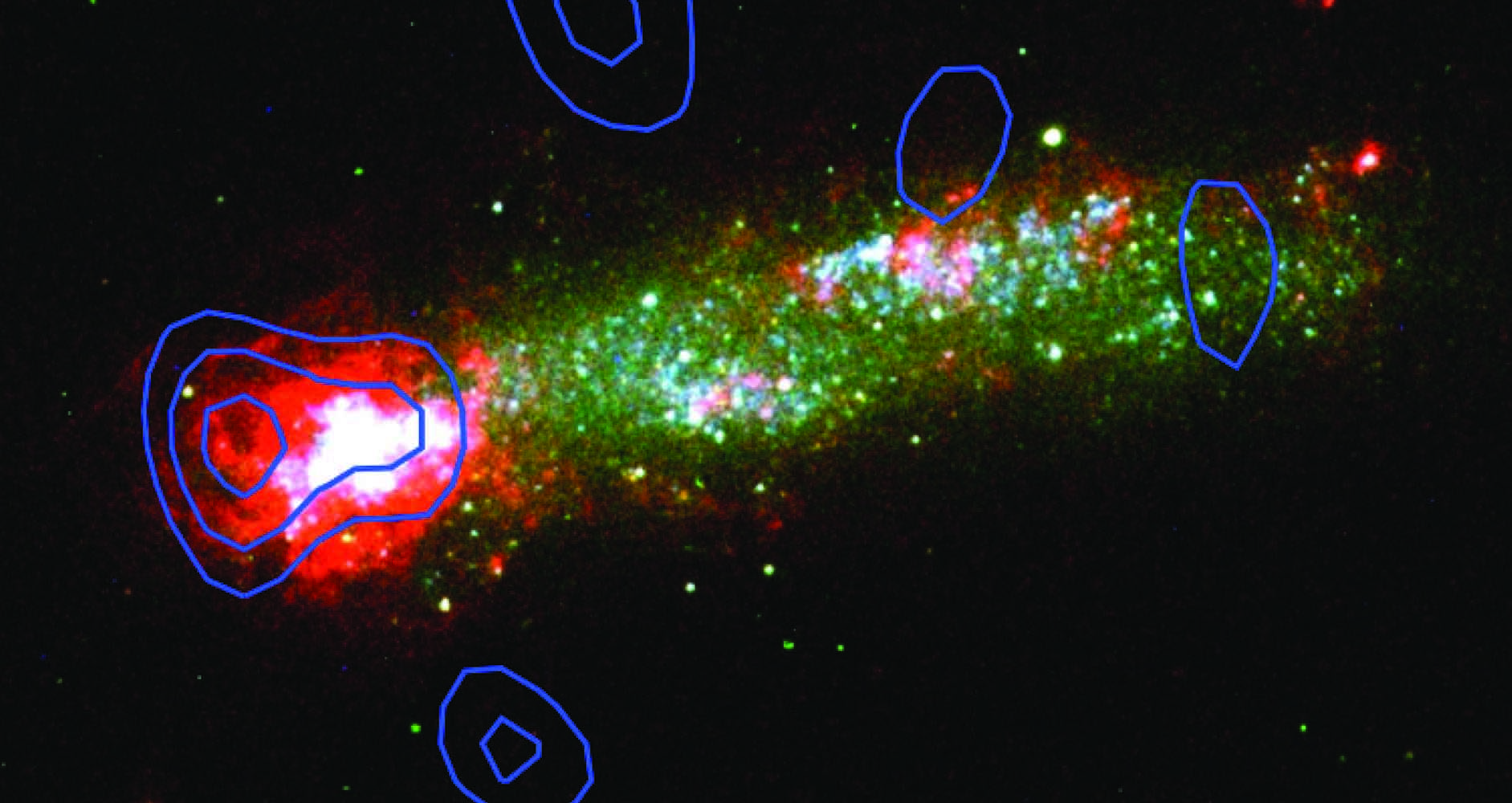}
\caption{Color image of Kiso 5639 from HST \citep{elmegreen16} in H$\alpha$,
V, B bands with superimposed CO contours at values of $2\sigma$, $3\sigma$ and $4\sigma$.
The main molecular cloud indicated by the contours contains $\sim3\times10^7\;M_\odot$ and
is offset from the bright star formation region near the head of this tadpole galaxy.
This lopsided structure is proposed to be the result of gaseous accretion or an impact event.
} \label{Fig_Kiso_HSTcolor_COcontour_cropped2}
\end{figure*}

\newpage
\begin{figure*}
\epsscale{1.15} \plotone{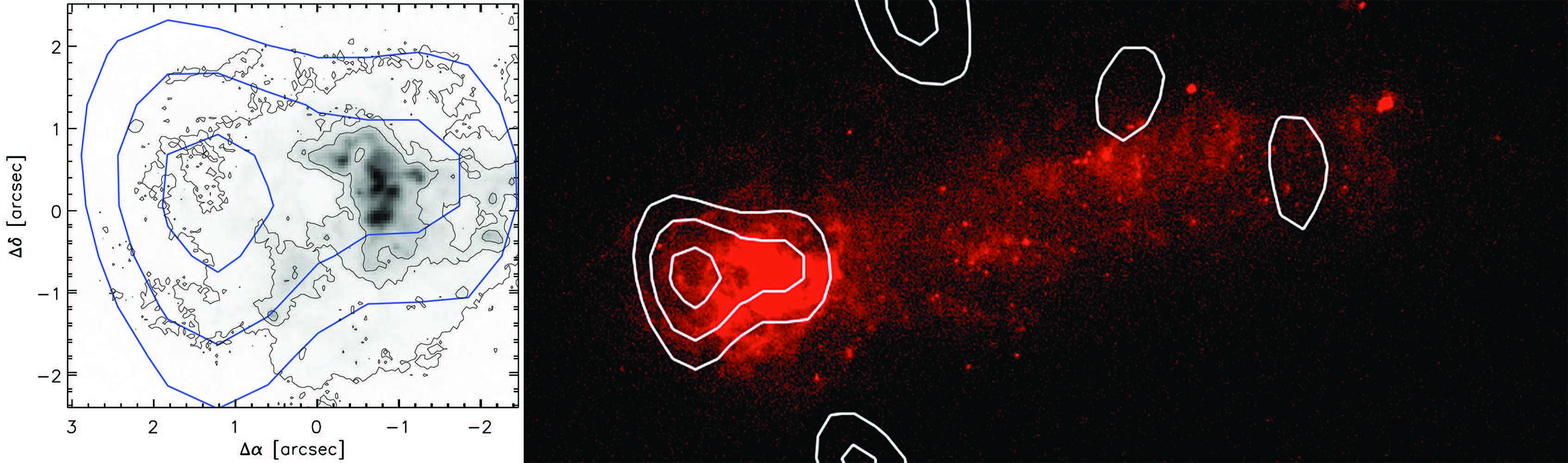}
\caption{The left image shows the H$\alpha$ emission in Kiso~5639 as a grayscale obtained with HST
\citep{elmegreen16} and enlarged around the head region near the giant molecular cloud.
The black contours are 3\%, 10\% and 20\% of the peak H$\alpha$ emission and the
blue contours are 2$\sigma$, 3$\sigma$ and 4$\sigma$
for the CO(1$-$0) integrated emission. Offsets positions are as in
Fig.~\ref{Fig1_CH_23apr2018}. The right image shows the HST H$\alpha$ and CO contours for the full galaxy.
} \label{Halpha_left_full_galaxy_right2}
\end{figure*}

\end{document}